\documentclass[aps,prb,groupedaddress, floatfix,showpacs, twocolumn]{revtex4-1}
	
\usepackage{graphicx,amsmath,appendix,simplewick,epstopdf,grffile,natbib}
\begin{document}

\title{Variational Monte Carlo study of spin polarization stability of fractional quantum Hall states against realistic effects in half-filled Landau levels}  
\author{ J. Biddle$^1$, Michael R. Peterson$^{2}$,  and S. Das Sarma$^1$} 
\affiliation{$^1$Condensed Matter Theory Center, Department of Physics, University of Maryland, College Park, Maryland 20742, USA}
\affiliation{$^2$Department of Physics \& Astronomy, California State University, Long Beach, California 90840, USA}
\date{\today}

\begin{abstract}
We compare ground state energies by variational Monte Carlo of the spin unpolarized Halperin 331 and the spin polarized Moore-Read (MR) Pfaffian fractional quantum Hall states at half-filling of the lowest Landau level (LLL) and the second Landau level (SLL) as a function of small deviations around the Coulomb point via the finite thickness effect and direct alterations to the the first two Haldane pseudopotentials.  In the comparison we find that in the LLL, either the 331 state or the MR Pfaffian may be lower in energy depending on the deviations.  In the SLL, however, the MR Pfaffian is consistently lower in energy except for  large deviations.  These results suggest that even under moderate deviations in the interaction potential (through various physical processes such as finite thickness, Landau level mixing, etc.), the MR Pfaffian description is more energetically favorable than the Halperin 331 state in the half-filled SLL (i.e. $\nu = 5/2$),  consistent with recent experimental investigations.
\end{abstract}
\pacs{73.43.-f, 71.10.Pm}

\maketitle
\section{Introduction}

The direct experimental observation of quasi-particles with non-Abelian anyonic statistics is an active area of research that may lead to the construction of a fault-tolerant topological quantum computer~\cite{DasSarmaTQC05,Nayak08}.  There are several condensed matter systems under study expected to exhibit non-Abelian excitations, including one-dimensional nano-wires adjoining s-wave superconductors~\cite{Lutchyn10}, cold atomic systems with spin-orbit interactions~\cite{Zhang08}, $p_x + ip_y$ superconductors~\cite{DasSarmaSC}, topological insulators on superconducting substrates~\cite{FuKane}, thin semiconductor films in contact with ordinary superconductors~\cite{Sau10} and others.  Perhaps the most discussed system where non-Abelian anyons are proposed to exist~\cite{MooreRead91,ReadGreen00,GreiterWenWilczek} is in the fractional quantum Hall effect~\cite{TSG1982} (FQHE) in the half-filled second Landau level~\cite{Willett87}, i.e., $\nu=5/2$, 
where strongly interacting electrons in two-dimensions (2D) are under the influence of a strong perpendicular magnetic field. Even though the FQHE at $\nu=5/2$ 
was first observed in 1987~\cite{Willett87} it remains as somewhat of an enigma since the physics of this FQH state differs considerably from the well-known Laughlin~\cite{Laughlin83} and composite fermion\cite{Jain89, jain2007composite} family of fractional quantum Hall states and is the only even denominator fraction to be observed in mono-layer quantum Hall systems so far.  At present, the leading theoretical description for $\nu = 5/2$, is the spin polarized Moore-Read (MR) Pfaffian wavefunction~\cite{MooreRead91} introduced in 1991, whose excitations have been theoretically shown to have non-Abelian anyonic statistics~\cite{MooreRead91,Nayak1996,BondersonGurarieNayak}.  Despite this prediction and considerable numerical evidence~\cite{Morf98,*StorniPRL} supporting the MR Pfaffian description, non-Abelian anyons have yet to be directly observed experimentally in quantum Hall systems, although there are recent tantalizing experimental results which are interpreted as being consistent with the 5/2 FQH quasiparticles being non-Abelian~\cite{WillettPNAS09,Willett09}.  But there exist alternative descriptions for the even-denominator FQHE that do not require non-Abelian statistics, and one of the leading alternatives is the unpolarized Halperin 331 state~\cite{Halperin}.  Part of the Halperin (m,m,n) family of Abelian fractional quantum Hall states~\cite{Halperin}, the 331 state is observed in half-filled quantum hall bi-layers where even denominator fractions are commonly observed and well understood as belonging to 331 type multicomponent Abelian even-denominator FQHE states~\cite{Suen92, Eisen92, SongHe93,Peterson2010a,Peterson2010b}.  Resolving the true nature of the $\nu=5/2$ (i.e. whether the excitations are Abelian or non-Abelian) would help determine if the fractional quantum Hall effect can potentially serve as a testbed for topological quantum computing.  For example, if the 331 type state can be convincingly ruled out for the 5/2 FQHE, this will provide further indirect evidence for the 5/2 FQHE as being the non-Abelian Moore-Read type state.

Determining the spin order (i.e., whether the actual 5/2 FQH state is spin polarized or not) could 
possibly help resolve the nature of the $\nu=5/2$ state.  Since the MR Pfaffian (Halperin 331) is, by construction, 
spin polarized~\cite{footnote1} (unpolarized), determining the spin polarization of the state can unequivocally rule-out either the MR Pfaffian or the unpolarized 331 description for $\nu = 5/2$.  
However the spin-order of $\nu = 5/2$ has a long history of inconclusive and seemingly contradictory studies.  One of the earliest experiments on this matter, for example, reported the collapse of the $\nu = 5/2$ FQHE state in a tilted magnetic field, suggesting an unpolarized FQHE state\cite{Eisenstein88}.   But this experiment was followed by a seminal numerical study by Morf\cite{Morf98} which showed that even in the limit of zero Zeeman energy, a polarized state has lower ground state energy per particle than a fully unpolarized state at $\nu = 5/2$.  Subsequent theoretical studies reported similar results\cite{Dimov, Feiguin}, and it is now understood that the collapse in the original ``tilt'' experiments was likely due to the appearance of  compressible 
``striped'' phases~\cite{Rezayi2000} or from orbital coupling of the in-plane magnetic field\cite{Peterson08}.  To avoid these issues, Liu et al.~\cite{Liu} performed an  ``altered'' tilt experiment where the in-plane magnetic field is kept small by varying the charge density and showed that the $\nu = 5/2$ state remains stable for relatively large tilt angles.  Recent rather impressive experiments performed by Tiemann et al.~\cite{Tiemann17022012} and Stern et al.~\cite{Stern12} provide strong evidence supporting a fully spin polarized 5/2 FQHE state.  But a notable exception is an earlier experiment by Stern et al.~\cite{Stern10} using photoluminescence spectroscopy to probe polarization, suggesting that the $\nu = 5/2$ state is actually unpolarized.  A similar result was also seen in a study by Rhone et al.~\cite{Rhone2011} where spin-order is probed via resonant light scattering.  It is possible that the signatures seen in these optical experiments~\cite{Stern10, Rhone2011} are due to local spin order near a charge impurity\cite{JainView,Wojs} and are, therefore, inconclusive, but such results still leave some doubt to the actual nature of spin-order in this state~\cite{DasSarmaFQHE10}.  The most convincing measurement to date are those by Tiemann et al.~\cite{Tiemann17022012} and Stern et al.,~\cite{Stern10}, which do indicate a spin-polarized 5/2 FQHE.  Our goal in the current work is to provide a reasonably complete study of spin-polarization comparing the candidate states MR Pfaffian versus 331 with respect to the 5/2 FQHE using direct numerical techniques in reasonably realistic theoretical models to see if one or the other can be ruled out purely on the basis of numerical studies.

An alternative explanation that may resolve the seemingly conflicting results on spin-order that we  explore here is the possibility that the $\nu = 5/2$ FQHE state is not unique\cite{JainView}.  In other words there may be more than one incompressible FQHE state, each with different spin polarization, that satisfy conditions to be experimentally observed at $\nu = 5/2$.  Whichever FQHE state is observed in an experimental sample may depend on certain details of the sample or experimental parameters because the competing polarized and unpolarized states may be relatively close in energy so that one or the other gets stabilized by the minute details of the sample parameters.  

In this work, we  explore this possibility by comparing the energy of two trial FQHE wavefunctions with respect to small deviations in the system Hamiltonian: the spin polarized Moore Read (MR) Pfaffian and the unpolarized Halperin 331 state. To examine the possibility of a phase change between spin polarized and unpolarized incompressible FQHE states within the $\nu = 5/2$ plateau, we focus entirely on the spin polarized MR Pfaffian and the spin unpolarized Halperin 331 state and examine how their respective ground state energies change with respect to small alterations to the effective interaction potential.  By 
appealing to the variational theorem, whichever state has the lowest ground state energy (strictly speaking, the lowest free energy) provides 
a better physical description--of course, the variational theorem cannot rule out the possibility of a lower energy 
ground state that we are not investigating (in fact, this is the likely scenario in the nu=1/2 state in the lowest Landau level where experimentally an FQHE has never been observed indicating that some kind of a compressible non-FQH state is likely to be lower in energy than either the MR Pfaffian or the Halperin 331 state).  To achieve this goal we alter the 2D Coulomb potential in two ways: i) through the finite thickness effect, and ii) by directly perturbing the first two Haldane pseudopotentials~\cite{Haldane83} (see below).  

In the finite thickness effect, the non-zero thickness of the quasi-2D electron system provides an effective potential slightly modified from the purely ideal 2D Coulomb potential, which we refer to as the ``Coulomb point" in this work.  Since FQHE states have been shown to be sensitive to this finite thickness effect and the thickness is expected to vary for different experimental samples, the finite thickness effect is a natural area to investigate~\cite{PetersonPRL08,Peterson08} (in these references the 
finite-thickness effect was considered in the context of spin polarized electrons via exact diagonalization).  Although our results do not suggest a direct quantum phase transition between the MR Pfaffian and the Halperin 331 in the second Landau Level (SLL) induced by tuning the finite thickness, the energy difference between the two states decreases with sample thickness, implying that other perturbations may make either state energetically favorable for very deep wells; some of these other 
perturbations could be Landau level mixing, disorder, effect of nearby gates, self-consistency of the confining potential itself, etc., which are not considered in our work since they are far too sample-specific to be treated theoretically at this stage..  

Our second approach to altering the 2D Coulomb potential -- directly perturbing the first two Haldane pseudopotentials~\cite{Haldane83} -- provides us a general theoretical probe that can identify areas of interest which may be reached experimentally via other effects.  The Haldane pseudopotentials, $V_m$, parametrize the effective interaction potential in terms of the relative angular momenta $m$ between two particles, and thus, perturbing the first two pseudopotentials (i.e. $m$ = 1,2) alters the short range interactions between electrons (note that in this study, we leave the $m$ = 0 term fixed).  The method of altering 
pseudopoentials is a common approach taken in FQHE exact diagonalization numerical studies aimed at probing sensitivities to different moments in the interaction strength, however, this approach has not yet been attempted in VMC studies to the best of our knowledge.   We note that our two alternative ways of introducing 'small deviations' or tuning away from the pure Coulomb point (realistic finite thickness effect and varying the lowest pseudopotentials) are complementary theoretical methods of tuning the system Hamiltonian since the finite thickness correction modifies all the Coulomb  pseudopotentials in a complex manner which cannot simply be simulated by changing the two lowest pseudopotentials.

We also examine the lowest Landau level (LLL) (i.e. $\nu = 1/2$) in addition to the second Landau level (SLL) (i.e. $\nu = 5/2$) for comparison.  In the LLL, no incompressible even-denominator FQHE has been experimentally observed in mono-layer systems to date, but there are several theoretical proposals to engineer certain experimental conditions in such a way that even-denominator states are energetically favorable~\cite{Scarola10}.  In these cases the Halperin 331 state and the MR Pfaffian are likely possibilities.  In our study of the LLL, we find that either state may be energetically favorable in the LLL depending on the pseudopotential deviations.  This is in contrast to our results in the SLL where we find that the MR Pfaffian is generally lower in energy than the Halperin 331 state for most deviations examined, suggesting that the MR Pfaffian description is, indeed, better suited for the half-filled SLL, i.e., $\nu = 5/2$.  This is of course also consistent with the most recent experimental status of the subject where the SLL 5/2 FQHE appears to be spin-polarized.  Our work, however, indicates that the corresponding LLL situation is more delicate, and if an incompressible FQHE is ever observed in a half-filled LLL monolayer 2D system, it could either be a MR spin-polarized Pfaffian or a Halperin spin-unpolarized 331 state.

We add a theoretical subtlety here which has sometimes caused some confusion in the literature.  The Halperin 331 state in general does not obey the full SU(2) symmetry (specifically, the so-called Fock condition necessary for a spin-independent many-body Hamiltonian which must conserve the total spin of the system), and cannot therefore be a true eigenstate of the single-layer Coulomb Hamiltonian since by definition this Hamiltonian obeys the full SU(2) symmetry because the Coulomb interaction is spin-independent--the 331 state was originally conceived for the double-layer 2D system where the Coulomb interaction does indeed depend on the layer index and is in general not SU(2) invariant in the layer index.  This is, however, not a problem for our VMC analysis since we are only interested in comparing energies between variational ground states (which do not care about the symmetry of the Hamiltonian) and are not trying to obtain the exact theoretical eigenstate of the system (which would be an impossible task any way, even the MR Pfaffian can at best be a good variational state for the system and by no means the exact eigenstate).  In the end, the best theory we can hope for is to obtain a variational ground state (MR Pfaffian or 331) which is adiabatically connected to the exact ground state of the experimental system without any intervening quantum phase transition so that the spin-polarization status of the variational  ground state and the exact ground state remains the same.  Thus, for our purpose, both the MR Pfaffian and the 331 are perfectly (and equally) legitimate variational choices, and which ever has lower VMC energy could be construed as the ``correct" ground state of the system (at least within the narrow, but very reasonable, restricted variational choice of only two candidate wavefunctions).

We also mention that all our work leaves out the trivial Zeeman energy of the system arising from the applied magnetic field creating the Landau levels in the first place, which helps the spin-polarized state over the spin-unpolarized state.  Since the applied field is typically rather small for the $\nu = 5/2$ FQHE, leaving out Zeeman energy (which is trivial to include for any given field) is probably a reasonable approximation, but it is helpful to remember that even if a spin-unpolarized ground state arises from our VMC analysis, the Zeeman energy could in principle eventually win over, leading to the experimental state being spin-polarized. The reverse, however, is not true, i.e. if the zero-Zeeman splitting situation has a (spontaneously symmetry-broken) spin-polarized ground state, it is unlikely that finite Zeeman splitting will change the ground state to a spin singlet.

\section{Variational Monte Carlo Evaluation of the Energies Using Effective Potentials} \label{sec:Method}

We use variational Monte Carlo (VMC) methods in the same spirit as Refs. \onlinecite{Park98} and \onlinecite{Dimov} to estimate the energy per particle of the Halperin 331 state and the MR Pfaffian state for altered Coulomb potentials in the lowest and second Landau level with up to $N=120$ electrons and extrapolate to the thermodynamic limit ($1/N \rightarrow 0$).  To examine the finite thickness effect, we use the potential derived for the infinite square well potential in a planar system which is given, in Fourier space, by \cite{DasSarma85}
\begin{eqnarray}
V_{\mathrm{SQ}}(k) &=& \frac{e^2}{\epsilon l}\frac{1}{k}\int dz_1 dz_2 |n(z_1)|^2|n(z_2)|^2e^{-k|z_1-z_2|} \nonumber \\
&=&\frac{e^2l}{\epsilon k}\frac{\left\{3kd + \frac{8\pi^2}{kd}- \frac{32\pi^4(1-e^{-kd})} {(kd)^2\left[(kd)^2+4\pi^2\right]}\right\}}{(kd)^2+4\pi^2}, \label{eq:Vwell}
\end{eqnarray}
where $l$ is the magnetic length given by $\sqrt{\hbar c/e B}$ and $d$ is the thickness.  There are other finite thickness potentials that we can explore~\cite{Tsuneya82, Stern84, DasSarma85, Zhang86}, however we expect the infinite square well potential to be sufficient in obtaining a qualitative comparison between the two states\cite{PetersonPRL08,Peterson08}.  

We also examine the effect of directly perturbing the Haldane pseudopotentials~\cite{Haldane83} $V_m$ for the Coulomb potential in LLL and SLL.  In particular, we examine the effect of the perturbations $\tilde{V}_1 \rightarrow V_1 + \Delta V_1$ and $\tilde{V}_2 \rightarrow V_2 + \Delta V_2$ for pseudopotentials derived from the Coulomb potential (i.e. $V(k) = 1/k$).  In the planar geometry, the Haldane pseudopotentials for the $n$th Landau level are given in terms of the Fourier transform of the effective interaction potential $V_{\mathrm{eff}}(k)$ by 
\begin{equation}
V_m^{(n)} = \int^{\infty}_{0}{dk k [L_{n}(k^2/2)]^2L_m(k^2)e^{-k^2}V_{\mathrm{eff}}(k)} \label{eq:pseudo}
\end{equation}
where $L_n(x)$ are Laguerre polynomials.  Note that we are using planar pseudopotentials due to two primary reasons:  (1) it is simpler to use 
planar $V_{m}^{(n)}$'s and (2) for such large systems $N\sim100$ examined in our studies there is very little difference between spherical 
and planar pseudopotentials.

\begin{figure}[h]
	\includegraphics[width=0.47\textwidth]{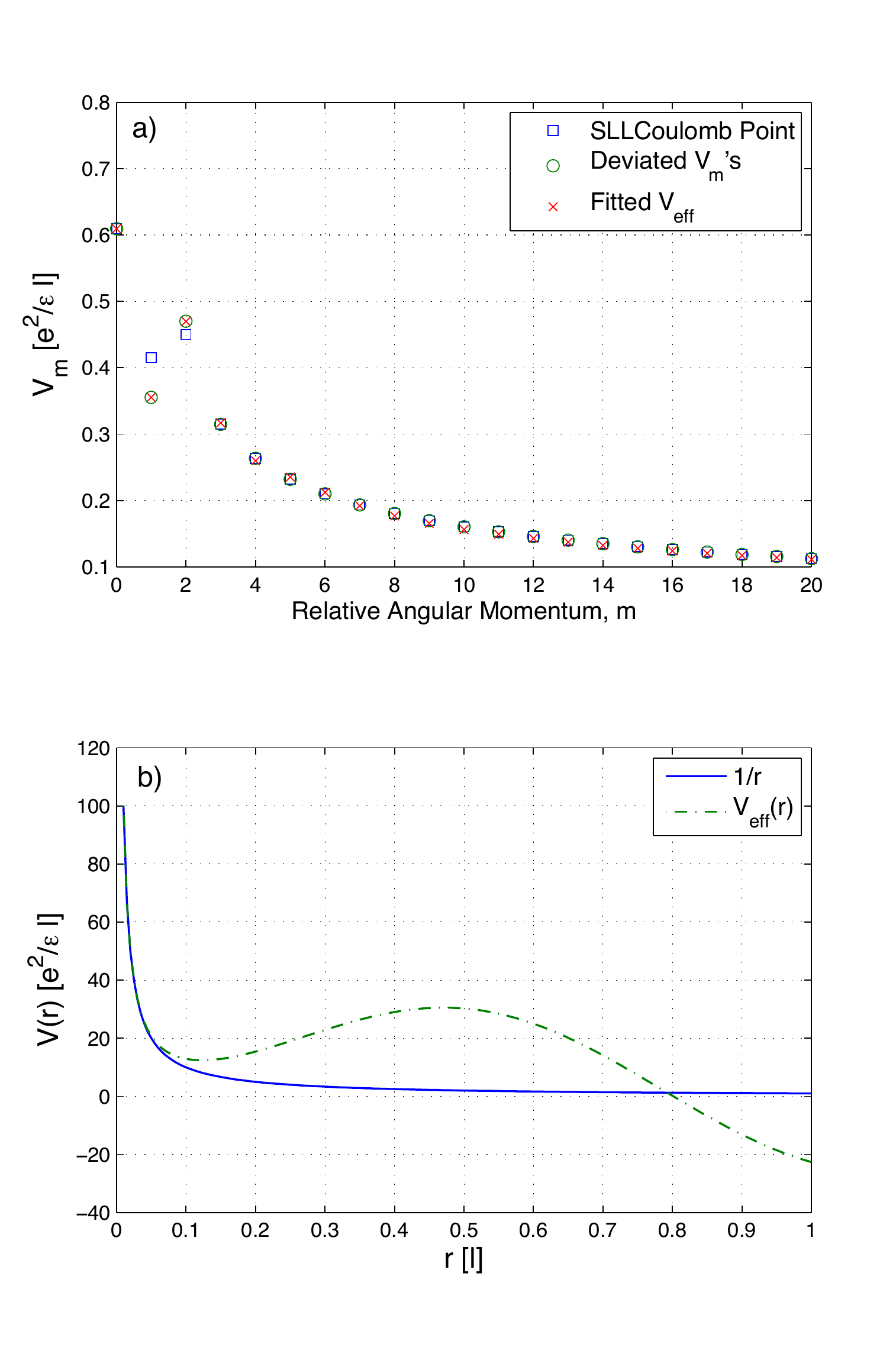}               
  \caption{a) Perturbed pseudopotentials $\tilde{V}_m$ for $\Delta V_1 = -0.06$ and $\Delta V_2 = 0.02$ in the SLL (the Coulomb point is the 
  Coulomb interaction in the SLL) and the corresponding fitted pseudopotentials from Eq.~(\ref{eq:veff}).   Note that any difference between the pseudopotentials calculated from the real space 
  effective potential corresponding to the deviated pseudopotentials and the deviated pseudopotentials themselves is smaller than the symbols on 
  the figure.  b) The resulting $V_{\mathrm{eff}}(r)$ compared to the Coulomb potential $V(r) = 1/r$}\label{fig:fit_example}
\end{figure}

In order to use VMC methods to estimate wavefunction energies, we require an effective potential in real space, $V_{\mathrm{eff}}(r)$, such that the application of Eq.~(\ref{eq:pseudo}) results in our perturbed pseudopotentials, $\tilde{V}_m$, on the left-hand-side.  The immediate difficulty we run into is that there is no clear procedure to invert Eq.~(\ref{eq:pseudo}) to obtain $V_{\mathrm{eff}}(r)$ for arbitrary $\tilde{V}_m$--it is a rather non-unique
one-to-many mapping.  Also, even in the unperturbed case, estimating energies in the SLL is not straight-forward since most FQHE trial wavefunctions under study do not have a closed-form expression in the SLL.  To get around these difficulties, we chose a variable effective potential with fitting parameters, $c_i$, and set these parameters such that the 
result of applying Eq.~(\ref{eq:pseudo}) on the effective potential very closely matches the perturbed pseudopotentials.  And when examining the SLL, we ``simulate'' the SLL in the LLL by fitting the effective potential within the LLL to the perturbed SLL pseudopotentials~\cite{Park98}, that is, 
we project the SLL into the LLL.  Several forms for the effective potential have been used for previous Monte Carlo studies of the FQHE \cite{Park98, Scarola00,Toke05}.  For our study, we use the following form for the effective potential (in units of $e^2/\epsilon l$):      
\begin{equation}
V_{\mathrm{eff}}(r) = \frac{1}{r} + \sum_{i=1}^{M} {c_i r^{2i} e^{r^2}} \label{eq:veff}.
\end{equation}
We choose this form because for large enough $M$, the potential fits both even and odd pseudopotentials to a reasonable degree -- only odd  pseudopotentials are important when fully polarized or spinless wavefunctions are under investigation -- and the fits to $V_m$ for large $m$ (i.e. $m > M$) are generally consistent across different perturbations, $\Delta V_1$ and $\Delta V_2$, allowing us to make fair comparisons between different perturbations.  In choosing the number of terms, $M$, in the effective potential, there is a trade-off between tighter fits to the pseudopotentials for larger $M$ and ease with which the Monte Carlo converges -- the addition of terms in Eq.~(\ref{eq:veff}) leads to an oscillatory potential that takes, in general, more iterations to reach Monte Carlo convergence.  For our study, we use $M = 6$.  As an example, we show in Fig.~\ref{fig:fit_example} perturbed pseudopotentials $\tilde{V}_m$ for $\Delta V_1 = -0.06$ and $\Delta V_2 = 0.02$ in the SLL and the corresponding fitted pseudopotentials resulting from a non-linear least squares fit of Eq.~(\ref{eq:veff}) to $\tilde{V}_m$ via Eq.~(\ref{eq:pseudo}).  It is worth noting that  the $V_m$'s calculated 
from the effective potential for $m>M=6$ are very good approximations to the actual values and only differ at the level of a fraction of a percentage 
point ($\sim 0.6\%$ on average)

%%%%%%%%%%%%%%%%%%%%%%%%%%%%%%%%%%%
%%%%%%%%%%%%%%%%%%%%%%%%%%%%%%%%%%%%%%%%%%%%%%%%%%
Throughout this work we make use of the spherical geometry where electrons are confined to a 2D spherical surface of radius $R$ with a magnetic monopole of magnetic charge $Q$ at the center of the sphere\cite{Haldane83,Fano86}.  The radius of the sphere is determined by the magnetic charge $Q$: $R^2 = Q$.  The magnetic charge for a quantum Hall state with $N$ electrons at filling factor $\nu$ is given by $2Q = N/\nu + \chi$ where $\chi$ is the topological shift\cite{Wen95} and depends on the FQHE state under investigation.  Lastly, the distance between two electrons on the spherical surface 
is taken to be the cord distance.  

In the spherical geometry, the unpolarized Halperin 331 and polarized MR Pfaffian states are given by
\begin{eqnarray}
\label{eq:halsph}
\psi_{331}&=&\hat{\mathcal{A}}\prod_{i<j} {(u^{\uparrow}_{i} v^{\uparrow}_{j}-v^{\uparrow}_{i} u^{\uparrow}_{j})^3}\prod_{k<l} {(u^{\downarrow}_{k} v^{\downarrow}_{l}-v^{\downarrow}_{k} u^{\downarrow}_{l})^3} \nonumber \\ &&\times
\prod_{r,s} {(u^{\uparrow}_{r} v^{\downarrow}_{s}-v^{\uparrow}_{r} u^{\downarrow}_{s})}
\end{eqnarray}
and
\begin{equation}
\hspace{-2cm}\psi_{\mathrm{Pfaff}}=\prod_{i<j} {(u^\uparrow_i v^\uparrow_j - v^\uparrow_i u^\uparrow_j)^2} \mathrm{Pf}[M]
\label{eq:pfaffsph}
\end{equation}
where $u_i = \cos(\theta_i/2)\exp(i\phi_i/2)$ and $v_i = \sin(\theta_i/2)\exp(-i\phi_i/2) $, and $u^\sigma_i=u_i\otimes |\sigma\rangle$ and 
$v^\sigma_i=v_i\otimes |\sigma\rangle$, $|\sigma\rangle$ is the spin ket, $\hat{\mathcal{A}}$ is the antisymmetrization operator, and
 Pf[$M$] is the Pfaffian of the matrix $M_{i,j} = (u^\uparrow_i v^\uparrow_j - v^\uparrow_i u^\uparrow_j)^{-1}$.  
 The magnetic charge for both states is given by $2Q = 2N - 3$. 

To evaluate the energy of some wavefunction $\psi$ via variational Monte Carlo we calculate the Hamiltonian 
expectation value
\begin{eqnarray}
E_\psi &=& \frac{\int d\mathrm{\Omega}_1\ldots d\mathrm{\Omega}_N \psi^*(\mathrm{\Omega_1},\ldots,\mathrm{\Omega}_N)\hat{H}
\psi(\mathrm{\Omega_1},\ldots,\mathrm{\Omega}_N)}
{\int d\mathrm{\Omega}_1\ldots d\mathrm{\Omega}_N |\psi(\mathrm{\Omega_1},\ldots,\mathrm{\Omega}_N)|^2}\nonumber\\
&=&\frac{\int d\mathrm{\Omega}_1\ldots d\mathrm{\Omega}_N |\psi(\mathrm{\Omega_1},\ldots,\mathrm{\Omega}_N)|^2 \hat{H}}
{\int d\mathrm{\Omega}_1\ldots d\mathrm{\Omega}_N |\psi(\mathrm{\Omega_1},\ldots,\mathrm{\Omega}_N)|^2}
\end{eqnarray}
where $\mathrm{\Omega}=(\theta,\phi)$ is the particle position on the sphere and $\hat{H}=\sum_{i<j}^N V^\mathrm{eff}(|\mathrm{r}_i-\mathrm{r}_j|)$. 
When calculating the energy of $\psi_\mathrm{Pfaff}$ we make use of the identity, $\det M = \left|\mathrm{Pf}[M]\right|^2$.

Before presenting the effects of finite-thickness and directly perturbing Haldane pseudopotentials we briefly discuss the background energy.  It is assumed that there is a uniform distribution of positive charge on the spherical surface so that the total energy is negative and the electron's state represents a stable phase of matter.  That is, we place $N$ positive charges on the surface of the sphere and calculate the interaction energy between an electron and the 
background $E_{el-bg}$ and the interaction energy of the background with itself $E_{bg-bg}$.  For a pure Coulomb interaction this works out to be
\begin{eqnarray}
E_{el-bg}&=&-\frac{N^2 e^2}{\sqrt{Q}\epsilon l}\;,\\
E_{bg-bg}&=&\frac{N^2 e^2}{2\sqrt{Q}\epsilon l }\;,
\end{eqnarray}
yielding $E_{bg} = E_{el-bg}+E_{bg-bg}=-\frac{N^2 e^2}{2\sqrt{Q}\epsilon l}$.  Remember that the radius of the sphere is $R=\sqrt{Q}$.  
Now, strictly speaking, this energy comes about by doing a rather trivial integral over the surface of the sphere with the distance between particles 
defined as the cord distance instead of the arc distance--in the thermodynamic limit both choices are equivalent.  

For our calculations it is a little bit more subtle.  We are considering electrons projected into the LLL with 
effective potentials that take into account finite thickness, electrons completely 
confined to the SLL, and potentials produced by small deviations away from the Coulomb point through the direct manipulation of $V_1$ and $V_2$.  
We find this effective potential through Eq.~(\ref{eq:veff}) and to get the proper background energy we calculate
\begin{eqnarray}
E_{el-bg} &=& -\frac{e^2N^2}{2}\int_0^{\pi} d\theta \sin(\theta) V_{\mathrm{eff}}(r(\theta))\;,\\
E_{bg-bg} &=& -\frac{1}{2}E_{el-bg}\;,
\end{eqnarray}
where $r(\theta) = 2R\sin(\theta/2)$.  While there is no deep physics hidden in the background energy, it is needed to ensure that the ground state energy per particle has a well-defined and 
finite thermodynamic limit.  Further, since we are comparing two ground state energies, this background energy cancels out in a sense.

\section{Ground state energies for effective potentials in the LLL and SLL} \label{sec:Results}

%%%%%%%%%%%%%%%%%%%%%%%%%%%%%%%%%%%%%%%%%%%%%%%%%%%%%%
\begin{figure}
	\includegraphics[width=0.47\textwidth]{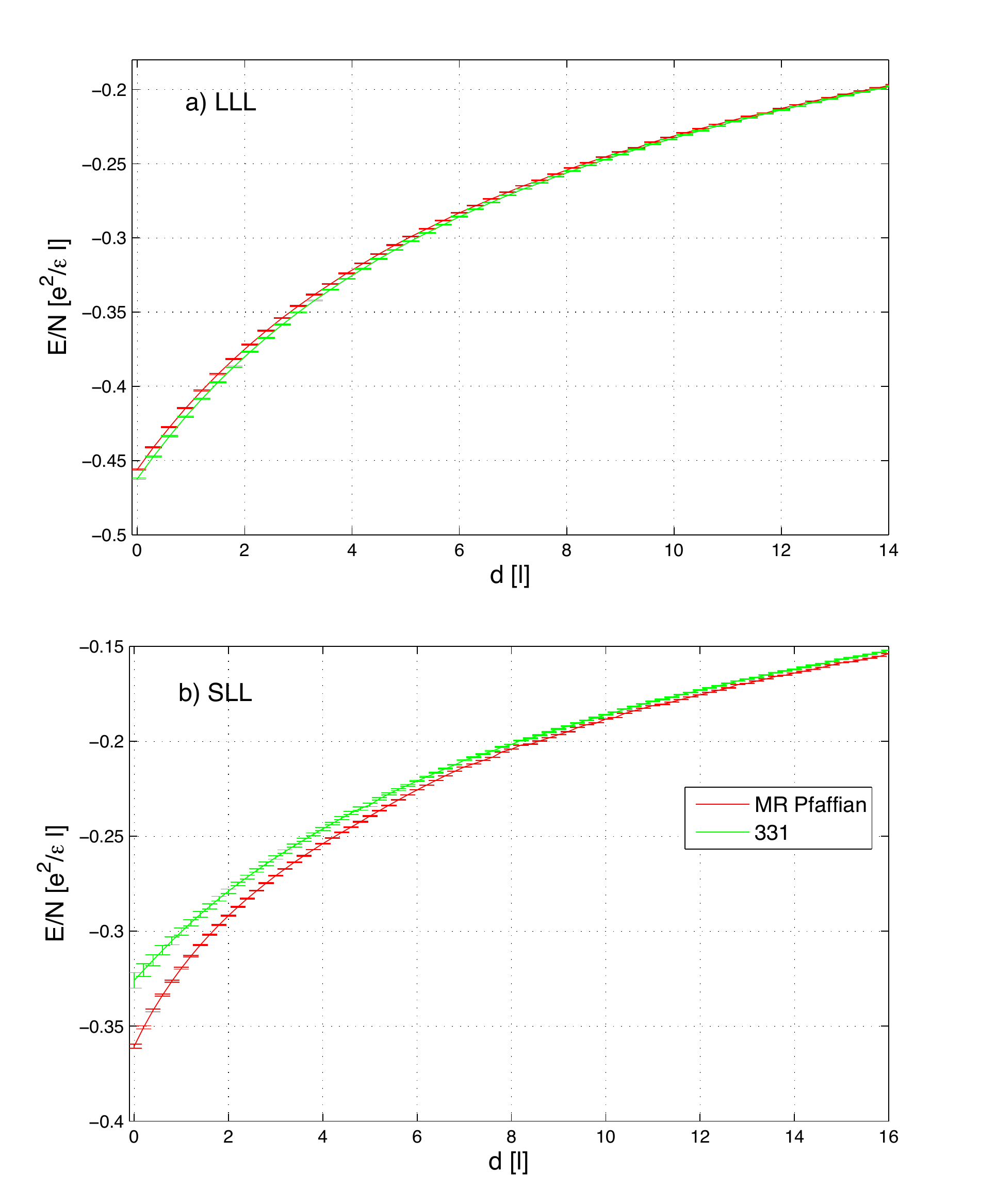}               
  \caption{ Energy per particle (in units of $e^2/\epsilon l$) as a function of finite thickness, $d$ (in units of the magnetic length, $l$) for the MR Pfaffian (red) and the Halperin 331 state (green) in the LLL (a) and the SLL (b).  The energy per particle increases with thickness and the difference 
  in energy between the two states decreases.  However, in the LLL, the Halperin 331 is always lower in energy than the MR Pfaffian while, 
  in the SLL, the opposite is true.  In other words, finite thickness alone does not apparently drive a spin order transition. }\label{fig:FT}
\end{figure} 
%%%%%%%%%%%%%%%%%%%%%%%%%%%%%%%%%%%%%%%%%%%%%%%

%%%%%%%%%%%%%%%%%%%%%%%%%%%%%%%%%%%%%%%%%%%%%%%%%%
\begin{figure}
	\includegraphics[width=0.47\textwidth]{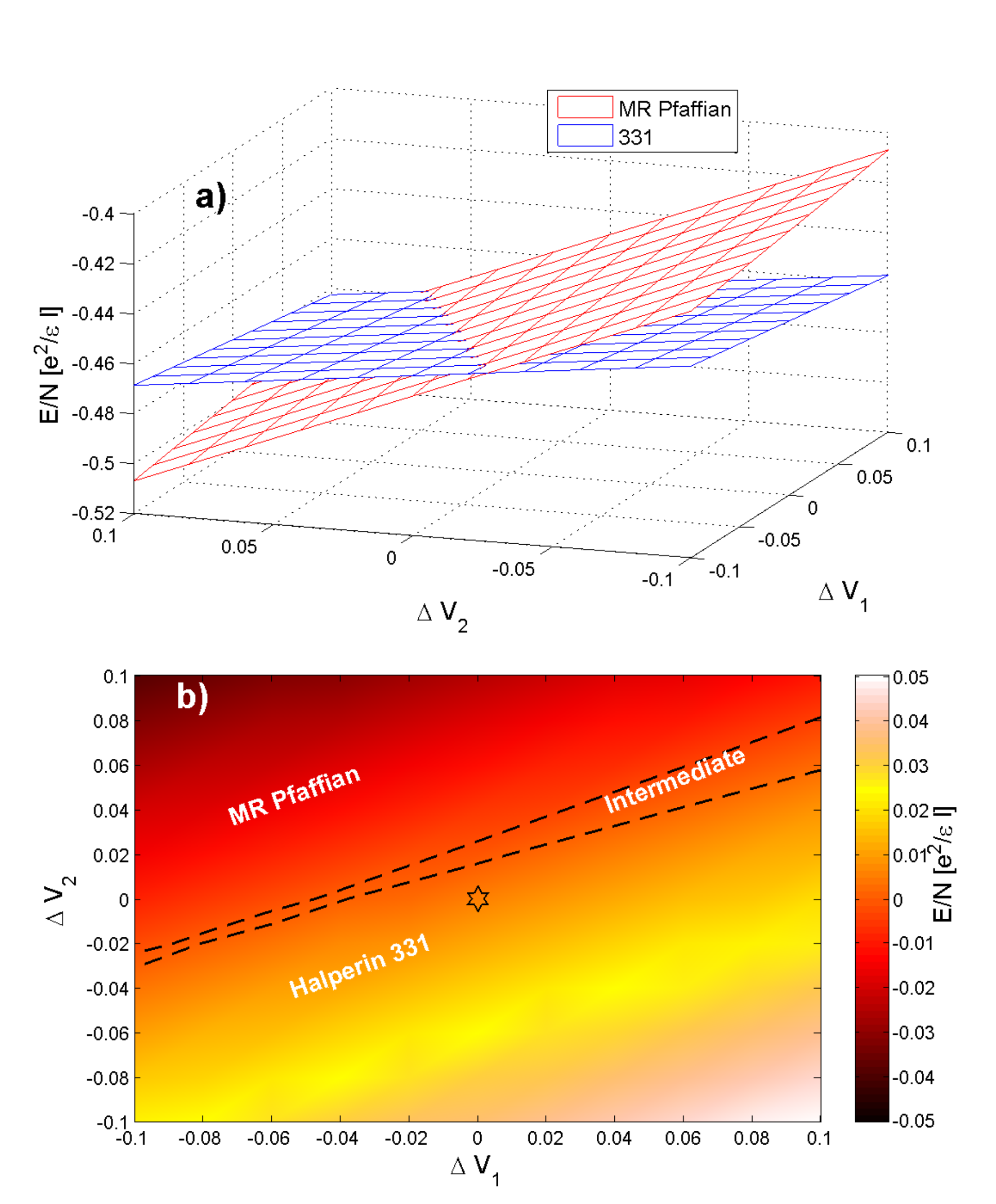} \\
	  \caption{ Comparison of Halperin 331 and MR Pfaffian in the LLL. a) Energy per particle as a function perturbation strengths $\Delta V_1$ and $\Delta V_2$.  b) Energy difference $E_{\Delta} = E_{\mathrm{Pfaff}} - E_{331}$ as function of $\Delta V_1$ and $\Delta V_2$.  Regions where either Halperin 331 ($E_{\Delta} > 0$) or MR Pfaffian ($E_{\Delta} < 0$) is energetically favorable are denoted.  The intermediate region denotes area where the energies are within numerical uncertainty of each other.  The star designates the Coulomb point for reference.  The statistical uncertainty 
	  in the energies is not indicated on these contour plots for ease of presentation.  However, it is similar in magnitude to 
	  what is presented in Fig.~\ref{fig:FT} but the qualitative effects of the uncertainty is indicated by the ``intermediate" regime where both energies are within 
	  statistical uncertainty of each other.   }\label{fig:LLL}
\end{figure} 
%%%%%%%%%%%%%%%%%%%%%%%%%%%%%%%%%%%%%%%%%%%%%%

%%%%%%%%%%%%%%%%%%%%%%%%%%%%%%%%%%%%%%%%
\begin{figure}
	\includegraphics[width=0.47\textwidth]{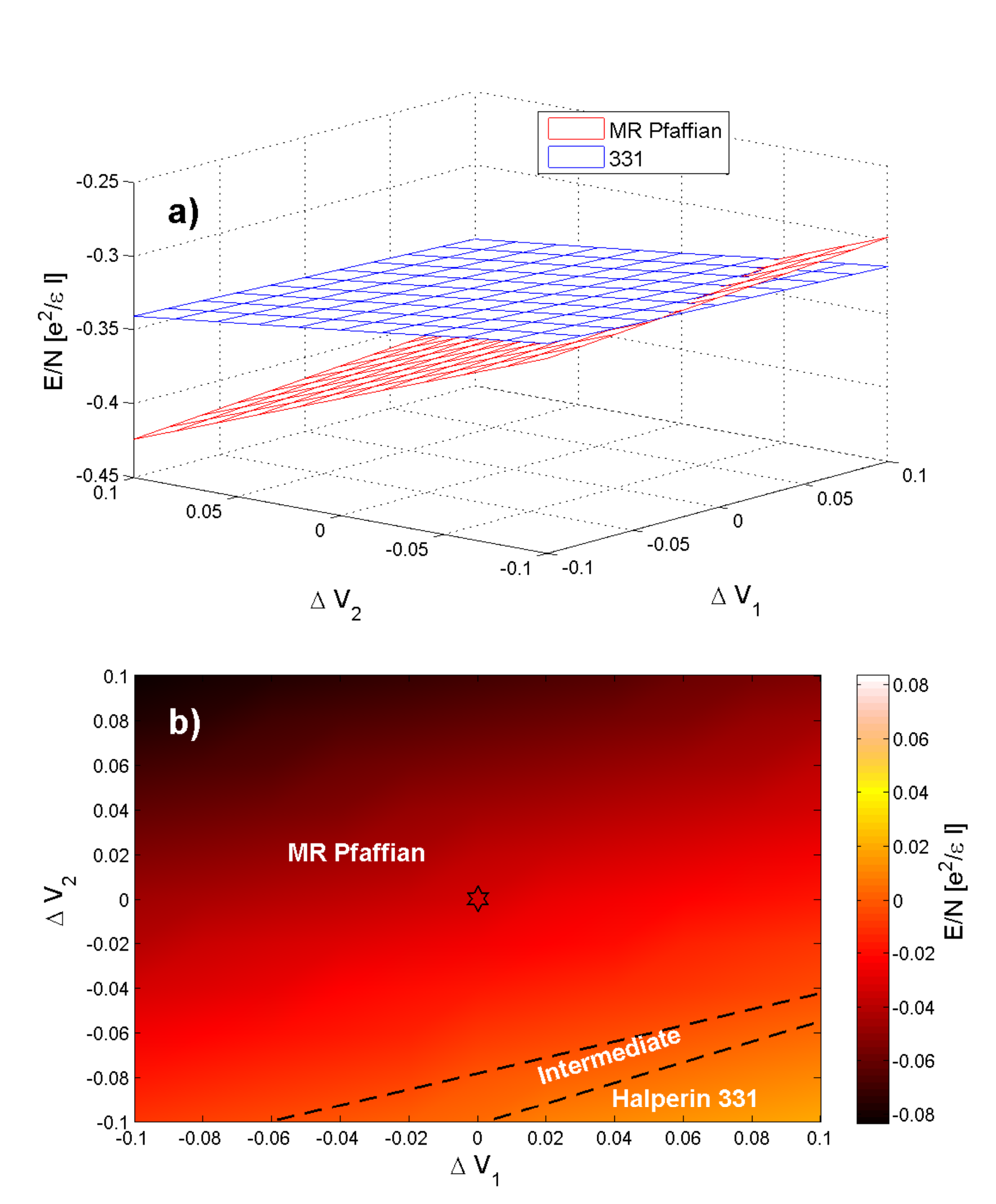}               
  \caption{ Comparison of Halperin 331 and MR Pfaffian in the SLL. a) Energy per particle as a function perturbation strengths $\Delta V_1$ and $\Delta V_2$.  b) Energy difference $E_{\Delta} = E_{\mathrm{Pfaff}} - E_{331}$ as function of $\Delta V_1$ and $\Delta V_2$.  Regions where either Halperin 331 ($E_{\Delta} > 0$) or MR Pfaffian ($E_{\Delta} < 0$) is energetically favorable are denoted.  The intermediate region denotes area where the energies are within numerical uncertainty of each other.  The star designates the Coulomb point for reference.  }\label{fig:SLL}
\end{figure} 
%%%%%%%%%%%%%%%%%%%%%%%%%%%%%%%%%%%%%%%%%%%%%
\begin{table}[t]
\caption{Energy calculated via VMC of various wavefunctions.  Our results for $E_{331}$ and $E_\mathrm{Pfaff}$ agree with those of 
Dimov \textit{et al.}~\cite{Dimov} at the Coulomb point in the LLL and SLL.  The results listed below for the polarized and unpolarized 
CFFS and the Haldane-Rezayi singlet state are given by Park \textit{et al.}~\cite{Park98}.  The lowest energy 
state at the Coulomb point in the LLL and SLL is the unpolarized CFFS and MR Pfaffian, respectively (both are indicated in bold).}
\begin{ruledtabular}
\label{table:vmc}
\begin{tabular}{lll}
Energy [$e^2/(\epsilon l)$] & LLL & SLL \\ \hline
$E_{331}$ 			  & -0.4631(3) 			        & -0.329(3) \\
$E_\mathrm{Pfaff}$  		  & -0.4573(3) 			        & {\bf-0.361(2)} \\
$ E_\mathrm{CFFS(P)}$      & -0.46557(6)~\cite{Park98}  & -0.3492(5)~\cite{Park98} \\
$E_{\mathrm{CFFS(UP)}}$  & {\bf-0.46953(7)}~\cite{Park98}  & -0.2952(3)~\cite{Park98}  \\
$E_\mathrm{HR}$ 		    & -0.3147(3)~\cite{Park98}    & -0.303(3)~\cite{Park98}    \\
\end{tabular}
\end{ruledtabular}
\end{table}

As a thorough numerical check, we first list our results for the MR Pfaffian and the Halperin 331 state at the Coulomb point with no perturbation in Table~\ref{table:vmc} and find 
that we are in agreement with Refs.~\onlinecite{Park98} and~\onlinecite{Dimov}.  
We note that other wavefunctions  are also possible, for example, a Composite Fermion fermi sea (CFFS), in a polarized (P) or 
unpolarized (UP) variety~\cite{Jain89,Halperin93,Rezayi2000} and the Haldane-Rezayi singlet state~\cite{Rezayi88} (HR), all whose 
energies in the LLL and SLL are listed in Table~\ref{table:vmc}.
In the LLL, it is clear that the lowest energy state is the unpolarized CFFS which is a gapless state that does not yield the FQHE, although, 
the energy of the 331 state is very close.  However, we know experimentally~\cite{Willett93} that no FQHE has yet been observed in single layer 
systems at $\nu=1/2$.  In contrast, at the Coulomb point in the SLL, the lowest energy state in Table~\ref{table:vmc} is the MR Pfaffian and, in fact, the MR Pfaffian 
has been routinely experimentally observed at $\nu=5/2$ albeit at low temperatures and in very high-quality samples, indicating that the $\nu=5/2$ FQHE is rather fragile with a very small gap with possibly  competing states with comparable energetics nearby in phase space.  

Since the purpose of this work is to investigate the spin polarization of the half-filled lowest and second Landau levels via VMC, we will focus exclusively 
on the Halperin 331 versus the MR Pfaffian wavefunctions.  A full investigation including all possible ansatz and more realistic effective 
potentials that include finite thickness \textit{and} Landau level mixing~\cite{Peterson13} is beyond the scope of this work and will have to await future works.  Our work is in the spirit of a restricted variational study which makes sense for this problem since the two  candidates we use (i.e. MR Pfaffian and Halperin 331)  are essentially the `only game in town' for incompressible even-denominator FQHE states in single-layer  systems.  In fact, part of the motivation of this work is to establish the feasibility of this sort of VMC investigation of Hamiltonians described by 
effective potentials.

Next, we examine the finite thickness effect and how it changes the expected ground state energy of the MR Pfaffian and the Halperin 331 state.  In Fig.~\ref{fig:FT}(a) we show the numerically calculated energies in units of $e^2/\epsilon l$, as a function of thickness (in units of the magnetic length, $l$) in the LLL.  The Halperin 331 state is consistently lower in energy but the gap between the energies of the MR Pfaffian and the Halperin 331 state decreases with increasing thickness.  Similar results are seen in Fig.~\ref{fig:FT}(b) for the SLL where, by contrast the MR Pfaffian is energetically favorable, but the energy difference between the two states again decreases with thickness.  Part of this likely stems from the fact that the overall energy scale is shrinking due to the finite thickness effect. We emphasize that the deceptive simplicity of Fig.~\ref{fig:FT} hides the complexity of the calculation.  For each point  on  both (rather the four) curves the following procedure was 
carried out:  (1) for each value of thickness $d$ the pseudopotentials were calculated, (2) a real space effective potential $V_\mathrm{eff}(r)$ was 
found from these pseudopotentials, (2) many   VMC evaluations of the 
energy of either $\Psi_{331}$ or $\Psi_\mathrm{Pfaff}$ for $N$ electrons were carried out, and finally (4)  these energies were extrapolated to the thermodynamic 
limit ($1/N\rightarrow 0$) to generate a single point.  The numerical work in producing Fig.~\ref{fig:FT} is very exhaustive.  

We now examine the effect of directly perturbing the first two Haldane pseudopotentials in the LLL and the SLL.  Fig.~\ref{fig:LLL}(a) gives the energy estimates for the MR Pfaffian and the Halperin 331 state with respect to $\Delta V_1$ and $\Delta V_2$ in the LLL.  Here we see that the energy of the MR Pfaffian is more sensitive to $\Delta V_m$ compared to the Halperin 331 state.  Also the MR Pfaffian decreases in energy with decreasing $\Delta V_1$ or increasing $\Delta V_2$, whereas the Halperin 331 energy shows opposing trends.  Figure \ref{fig:LLL} (b) shows the energy difference between the MR Pfaffian and the Halperin 331 state and an estimated phase diagram--the phase is determined by whichever 
wavefunction has the lowest ground state energy per particle.  The ``intermediate'' phase indicates where the energy differences of the two states are within numerical uncertainty of the VMC.  Here we see that the MR Pfaffian can be energetically favorable in the LLL for relatively small deviations from the Coulomb point for $\Delta V_2 > 0$ although the 331 state is lower in energy at the LLL Coulomb point. 

Results for perturbations about the SLL Coulomb point are given in Fig.~\ref{fig:SLL}(a).  Here we see the  energy versus $\Delta V_1$ and 
$\Delta V_2$ has similar trends as was found in the LLL , but the MR Pfaffian is consistently lower in energy in the SLL for the majority of perturbations under investigation.  Fig.~\ref{fig:SLL}(b) gives the difference in energy and an estimated phase boundary between the MR Pfaffian and the Halperin 331.  Again, the ``intermediate'' area indicates where the energies are within numerical uncertainty of each other.  Unlike in the LLL case, the MR Pfaffian is generally favored for any perturbation in the SLL.  In the region where the Halperin 331 state is favorable in the SLL, the perturbations result in $V_2 < V_1$, which is a qualitative feature of the Coulomb point in the LLL.  If the effect of $V_{m>2}$ are minimal, then we can argue that this region is qualitatively similar to the Coulomb point of the LLL and therefore, the Halperin 331 state is energetically favorable in this region given the results on the LLL.

\section{Conclusions} \label{sec:Conclusions}

Our results show that through the finite thickness effect, the energies of the Halperin 331 and the MR Pfaffian state increase with increasing sample thickness in either LL.  The energy difference between the two states decreases with increasing sample thickness, but there is no crossing between the two states in either LL, i.e., finite thickness apparently does not drive a spin polarization quantum phase transition at least for the situation with vanishing Zeeman energy considered in our work.  It is possible, in fact quite likely for the LLL, that a finite Zeeman splitting will induce a transition from the Halperin 331 to the MR Pfaffian state, but neither may be the true ground state in the LLL since no $\nu=1/2$ FQHE has ever been observed experimentally.  Additionally, our results show that the energy of the MR Pfaffian is more sensitive to changes in the pseudopotentials than the Halperin 331 state, where the MR Pfaffian energy decreases with increasing $\Delta V_2$  and decreasing $\Delta V_1$.  The energy of the Halperin 331 state, in contrast, is generally insensitive (in comparison to that of the MR Pfaffian) with slight decreases for $\Delta V_1 > 0$ and $\Delta V_2 < 0$, cf. Figs.~\ref{fig:LLL}(a) and~\ref{fig:SLL}(a).   In the LLL, the MR Pfaffian becomes energetically favorable for small increases in $V_1$ above the Coulomb point.  In the SLL, the Halperin 331 state becomes favorable for relatively large deviations from the Coulomb point, (i.e. $\Delta V_1 \gtrsim 0$ and $\Delta V_2 \lesssim -0.06$), again in the absence of the Zeeman energy -- given the small difference between the two VMC energies, it is quite likely that the MR Pfaffian state has lower energy than the Halperin 331 state for all thickness values and all deviations from the Coulomb point once the Zeeman energy is taken into account since the Zeeman energy would always prefer the spin-polarized state. 

Our study adds to the growing body of evidence\cite{Morf98,Feiguin,Stern12,Liu,Tiemann17022012} supporting the spin polarized MR Pfaffian description for FQHE at $\nu = 5/2$.  Of course, conclusive verification of the MR Pfaffian description requires the direct experimental observation of non-Abelian anyons.  But given the difficulty in conclusively detecting non-Abelian signatures~\cite{WillettPNAS09,Willett09, Kang}, novel experimental techniques will be needed for definitive verification.    

Lastly, we emphasize that our study additionally serves as a ``proof of principle" for VMC studies of various FQH systems that are described by effective 
potentials.  Effective potentials are needed when considering certain realistic effects such as finite thickness, Landau level mixing, higher Landau level 
FQHE, disorder, etc., or by simply artificially manipulating various Haldane pseudopotentials.  Our work establishes that the VMC technique is a viable alternative to the exact diagonalization method in theoretically studying the ground state properties of FQHE including various realistic effects which are often hard to implement using the finite size diagonalization method.  

\begin{acknowledgments}
This work is supported by Microsoft Q, NSF-JQI-PFC, and DARPA QUEST and M.R.P. is grateful for support through California State University 
Long Beach start-up funds.
\end{acknowledgments}


\begin{thebibliography}{58}
\bibitem[{foo()}]{footnote1}
\bibinfo{note}{While the MR Pfaffian wavefunction describes spin polarized
  electrons and exhibits non-abelian excitations, existence of non-abelian
  statistics does not strictly depend on full spin polarization-- a ground
  state with a slight deviation away from full spin polarization still of the
  ground state would still exhibit non-abelian excitations~\cite{ReadGreen00}.
  To the best of our knowledge it is not known precisely how large of a
  deviation away from full spin polarization is allowed but it is clear that
  fully unpolarized electrons in a half-filled Landau level would not support
  non-abelian anyons. Thus, in our study we ask the binary question of whether
  the ground state is fully spin polarized or fully unpolarized.}
\end{thebibliography}
\end{document}